%% file: main.tex
\documentclass{ptephy}

\usepackage{amsmath} 
\usepackage{graphics} 
\usepackage{url} 

\begin{document}

\newcommand{\gmtwo}{\ensuremath{g\!-\!2}}
\newcommand{\gmtwomu}{\ensuremath{(g\!-\!2)_{\mu}}}
\newcommand{\amu}{\ensuremath{a_{\mu}}}

\title{Measurement of muonium emission from silica aerogel }

\input{authors.tex}


\begin{abstract}%

  Emission of muonium ($\mu^{+}e^{-}$) atoms from silica aerogel
  into vacuum was observed.  Characteristics of muonium emission
  were established from silica aerogel samples with densities in
  the range from 29 mg~cm$^{-3}$ to 178 mg~cm$^{-3}$.  Spectra of
  muonium decay times correlated with distances from the aerogel
  surfaces, which are sensitive to the speed distributions,
  follow general features expected from a diffusion process,
  while small deviations from a simple room-temperature thermal
  diffusion model are identified.  The parameters of the
  diffusion process are deduced from the observed yields.

\end{abstract}

\subjectindex{C31, G04}

\parindent0pt

\maketitle

\section{Introduction} 

A new experiment has been proposed at J-PARC~\cite{J-PARC}
to measure the
anomalous magnetic moment $\amu = \gmtwomu/2$ of the muon. There
exists a discrepancy of about 3.5$\sigma$ between the best existing
measurement (0.54 ppm) \cite{Bennett2006} from
the Brookhaven experiment (E821) and the best theoretical
estimates \cite{HLMNT, Davier}. If the difference
persists with higher precision experiments and theoretical calculations, it
signals physics beyond the Standard Model. The Brookhaven method
will be used in a similar improved experiment at Fermilab, but the J-PARC
approach is quite different, and would be limited by much
different systematic uncertainties. The J-PARC experiment relies on the
acceleration of muons from essentially thermal energies, in order
to limit transverse momentum components and to enable their
eventual injection into a small storage device with a high-precision magnetic field. 
This requirement translates into the need for an ultra-slow muon source.

The most promising method to create an ion source for ultra-slow
muons is from laser ionization of thermal muonium in vacuum.
Muonium is the atomic bound state consisting of a positive muon
and an electron ($\mu^{+}e^{-}$ or Mu). It is chemically similar
to a hydrogen atom, but as an antilepton-lepton bound state it
has aspects in common also with positronium ($e^{+}e^{-}$ or Ps).
Muonium production occurs following thermalization of a positive muon
beam in a variety of gases, liquids, and solids.  Observation is
typically signified by the distinctive frequency of muonium
hyperfine transitions 
in weak transverse
magnetic fields (1.4 MHz at $10^{-4}$ T) that is observed in the
2.2\ $\mu$s decay time distribution of positrons following
parity-violating polarized $\mu^{+}\rightarrow
e^{+}\nu_{e}\bar{\nu}_{\mu}$ decays \cite{Brewer1975}.

Fundamental interactions of muonium in vacuum have been
investigated in several experiments such as searches for
conversion of muonium to antimuonium ($\mu^{-}e^{+}$)
\cite{Willman1999} and the precise measurement of the 1S-2S
energy interval in Mu \cite{Meyer2000,Chu1988}. Both required Mu
at low (near thermal) energies in an environment free of other
interactions, i.e., isolated in a vacuum.  A key to their success
was the development of techniques to produce Mu in a silicon
dioxide (silica) powder that subsequently emits it into a vacuum
\cite{Marshall1978,Beer1986,Woodle1988}.

Motivated by the extremely important \gmtwomu\ experiment, and by
its need for a reliable, intense, ultra-slow muon source, we have
undertaken systematic investigations of muonium emission into
vacuum from potential target materials. The silica powder of the
original method presented stability and handling problems, and
was considered incompatible with the high precision requirements
of \gmtwomu. Silica aerogel was identified as a promising
alternative, producing polarized muonium with reasonably high
probability from stopping muon beams, and also showing evidence of
muonium emission into vacuum. Unlike powder, it is
self-supporting, stable, and can be made in various sizes,
shapes, and densities. This paper presents the results of our
measurements of muonium behavior in a selection of high quality
silica aerogel samples.

\section{Aerogel samples} 

Four separate silica aerogel samples were investigated. All were
fabricated at Chiba University with the same method used for
high quality Cerenkov detector materials
\cite{Tabata2012}. Surfaces of all samples were prepared using a
process developed to make them hydrophobic, in order to limit
absorption of water following the supercritical drying of the
aerogels. The samples were manufactured in slabs of
$100\times100$~mm$^2$, but were cut to obtain a muon stopping
target of $30\times40$~mm$^2$, only slightly larger than the muon
beam transverse dimensions. The samples were prepared with
different densities of 29, 47, 97, and 178~mg\,cm$^{-3}$, and
with corresponding different thicknesses of 6.9, 4.8, 2.2, and
2.0~mm respectively. We wished to investigate the dependence of
muonium emission on density in the context of our diffusion
model. Higher density targets would provide higher proportions of
muons stopping near the surface of the target, thus potentially
leading to higher probability of muonium emission from that
surface. The reduction of thickness with higher density reduced
differences in target mass and consequently reduced
possible systematic variations due to different muon beam
requirements, although this was not practical for the highest density sample.


\section{Measurement}

\subsection{MuSR}

Based on our experience with muonium emission from silica powder,
the identification of silica aerogel from a longer list of
potential alternative candidate material samples was accomplished
using the method of muonium spin rotation and relaxation (MuSR).
The key properties that could be identified were: (1) a high
probability for polarized muonium formation, similar to powder
($0.64\pm0.03$ \cite{Kiefl1979}), (2) low depolarization, or long
persistence of the MuSR signal during the muon lifetime, and (3)
an indication that Mu exits the fundamental structures of the
material, signified by a distinctive increase in depolarization
of the MuSR signal when oxygen is present in the target. The
latter situation results from spin exchange interactions of
muonium with paramagnetic oxygen molecules, which have been
observed using gas \cite{Senba1989} and silica powder
\cite{Marshall1978,Kempton1990} as muonium-producing moderators.

In addition to the aerogel slab samples as described in the
previous section, the list of materials investigated included
porous forms of silica and alumina, commercially available
particulate aerogel (Nanogel) and silica powder (Cab-O-Sil) from
Cabot Corporation \cite{Cabot}, and a silica plate for calibrations and
comparisons. Porous silica showed a relatively small MuSR muonium
formation probability
($0.21\pm0.01$), but it was unaffected by the introduction of
oxygen, while alumina showed no significant signal even in vacuum.
Silica powder and particulate aerogel samples both produced
significant polarized Mu that showed depolarization with the
introduction of oxygen, but because they are not self-supporting
they were not investigated further except as a demonstration of
the oxygen depolarization technique.

\begin{figure}[t]
\begin{center}
\includegraphics[width=\textwidth]{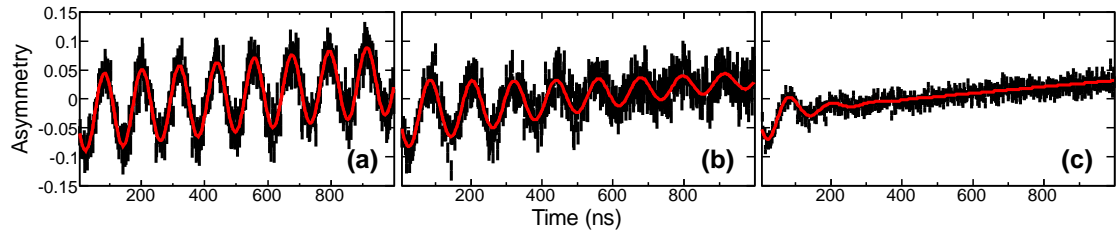}
\caption{\label{F:asy_plots}
Muonium spin rotation asymmetry distributions and fits. The muon
exponential decay time dependence has been removed. The muonium
signal relaxation rate is proportional to the concentration of oxygen in the aerogel
environment: (a) vacuum, (b) $0.50(1) \times 10^{16}$ mol
cm$^{-3}$ , and (c) $4.57(3) \times 10^{16}$ mol cm$^{-3}$.
}
\end{center}
\end{figure}

Based on muonium and muon asymmetries in MuSR measurements, the
fraction of muonium formed in the three aerogel samples was
$0.52\pm0.01$.
The MuSR relaxation rate was of order $0.05\ \mu\textrm{s}^{-1}$,
indicating long-lived muonium polarization. The relaxation rate
increased with the addition of oxygen into the sample environment
at a rate proportional to the oxygen concentration and
approximately consistent with comparable results for muonium in
gas moderators \cite{Senba1989} as well as for earlier silica
powder moderators \cite{Marshall1978}. The decay time
distributions were fit to oscillations of polarized muonium in an
applied transverse magnetic field of 0.6 mT, as shown in Fig.
\ref{F:asy_plots}.  The apparent non-flat background is due to
the much slower muon spin rotation signal.
For high oxygen concentrations, the signal disappeared very
quickly, indicating that nearly 100\% of muonium in these aerogel
samples was interacting with oxygen in the voids between the
aerogel structure prior to muonium decay.

Note that while observation of this oxygen interaction is
considered necessary for emission of muonium into a vacuum region
separated from the moderating material, it is not sufficient.
Useful production of muonium in vacuum for \gmtwomu\ must be
established by identification of Mu in a region where it could be
ionized by lasers, for example by the study of spatial and time
distributions of muon (muonium) decay.

\subsection{Space-time distribution of Mu}

\subsubsection{Experimental Setup} 

\begin{figure}[t]
\begin{center}
\includegraphics[width=0.9\textwidth]
{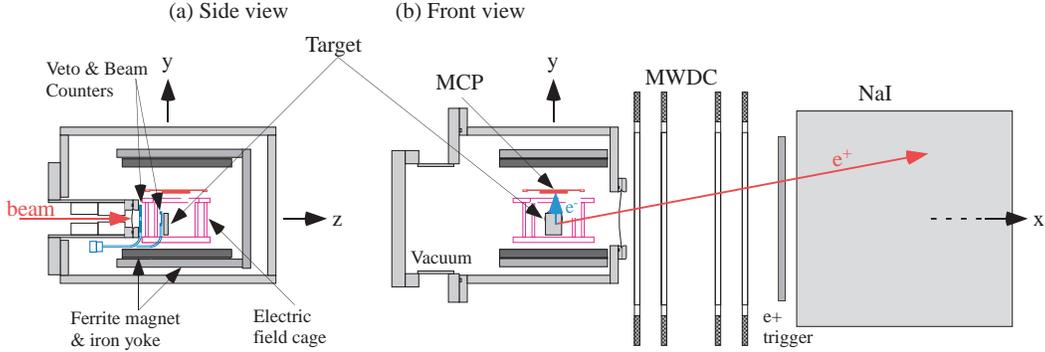}
\caption{\label{F:setup}
Setup for the muonium imaging measurement at the TRIUMF M15 beamline.
The axes of the coordinate system ($x, y, z$) are indicated in the figure.
}
\end{center}
\end{figure}

In order to measure the spatial and time distribution of muon
decay, a dedicated setup (Fig.~\ref{F:setup}) was built at the end of the
M15 surface muon channel at TRIUMF.  Here the coordinate system
($x$,$y$,$z$) is defined as follows.  The $z$-axis is along the
beam center with the origin $z = 0$ at the downstream surface of
the target, $y$ is in the vertical direction, and $x$ is
horizontal in the direction of the positron detection system. The
measurement principle basically follows the previous experiments
for silica powders~\cite{Beer1986,Woodle1988,Janissen1990}, where
the positron track was traced back to the plane bisecting the
sample to calculate the position of muon decay.

The selected aerogel sample was placed in the vacuum chamber 
with the longer dimension vertical and
with its downstream surface location fixed 
irrespective of the sample thickness. 
Muons passed the 10~mm hole in the veto scintillator and 
then through the 300~$\mu$m thick beam scintillation counter 
before entering the target.
Muons of the lowest practical momentum, typically $\sim$23 MeV/c
(``subsurface'' muons), therefore with the lowest spread $\Delta
p$ in momentum for fixed $\Delta p/p$, were used to increase the
fraction of the muon beam that stops in a thin layer, thus
reducing background from muons stopping elsewhere. 
For each
target, a scan of beam momentum was performed. For each momentum
in the scan, the number of muon decays observed from the target
was normalized to the number of muons entering the target region.
After corrections for dead time and rate effects, the result was
compared with a simulation to obtain the fraction of the beam stopping in
the target as the central momentum was varied. A flat-topped
Gaussian-smeared beam momentum distribution was used as input, where the
flat top width and the smearing width at low and high momentum
ends were set common to all samples.
The best values of the flat top width and smearing width 
describing the momentum scan data were 
5\% and 1.5\%, respectively, relative to the central momentum.
The central beam
momentum was obtained independently for each target.  This
allowed a test of the simulated beam properties, and provided
constraints for muon beam stopping distributions in each target.
The beam momentum was then adjusted for each target so that
approximately half of the incoming muons were stopped in the
target, while the remainder passed through mostly to regions
beyond those of interest for muonium decay in vacuum.
This adjustment served to maximize the muon stopping
density near the target surface at the downstream edge, nearest
the vacuum decay region. Due to the similarity of the aerogel target masses,
only small changes were necessary to find the optimum
momenta (22.9, 23.2, 23.1 and 24.5 MeV/c, respectively).
Thus systematic effects due to differences in muon
momentum were significantly reduced.

Multi-wire drift chambers (MWDC) and a NaI detector 
were used to register positrons from muon decay
and a micro-channel plate (MCP) detector was used to register
electrons following muon decay in the muonium.
The MWDC planes and readout electronics were previously used in
the TWIST experiment at TRIUMF~\cite{TWIST_MWDC}.
Four pairs of wire planes were used, with perpendicular planes in
each pair. Each plane was 4~mm thick, with 80 wires at 4~mm intervals.
The coordinates of hit wires were extrapolated to the plane
containing the beam and target axis to provide a
two-dimensional distribution of muon decay locations in and near
the target.
Since the thermal velocity of muonium in vacuum is of the order of 
5 mm/$\mu$s and the time duration is of the order of the muon lifetime
(2.2 $\mu$s), we needed a positional resolution of a few mm or less. 
The intrinsic positional resolution of the MWDC is 0.2~mm or better,
however the multiple scattering of positrons and the parallax
effect of the extrapolation actually limited the tracking resolution. 
The NaI detector was used to select only high energy positrons ($>$ 30~MeV/c),
reducing the effect of scattering.

%

Although the position sensitive multichannel plate (RoentDek HEX40 MCP) used to detect
the low energy electrons that remain when the muon in muonium decays did sharpen 
the event-by-event position of the muonium decays (and distinguish events in 
vacuum from those in the aerogel), we have not yet determined the 
efficiency uncertainties of the MCP, precluding its use in this quantitative analysis. 


\subsubsection{Extraction of space-time distribution}

A positron track was reconstructed by fitting a straight line to
hit coordinates in the MWDCs.  The confidence level of the fit was
required to be greater than 95\% to ensure good quality of
reconstruction.  A fiducial region was defined so that the positron
track projection to $x=0$\ lay within $\pm20$~mm in the vertical direction.
A cut on track slope, $|\frac{dz}{dx}|<0.1$, was applied to reduce the
parallax uncertainty in the extrapolation. Because low energy positrons
show larger multiple scattering angles, they were rejected
by requiring deposited energy in the NaI to be greater
than 30~MeV.

Reconstructed tracks were extrapolated back to the target area ($x=0$~mm)
with an accuracy estimated as 2~mm ($\sigma$)
by a GEANT-based Monte Carlo simulation~\cite{web:G4} .
This was confirmed using the experimental data taken with a solid silica
plate of 0.1~mm thickness that was used for calibration and
background estimation.

The time of muon decay was measured as the time of positron
detection in a pair of scintillators located behind the MWDCs,
with respect to the beam counter.  Corrections for
slewing effects and light propagation in the scintillator were
applied by using information on the pulse height of hits and
reconstructed track position on the scintillator. 
The time resolution was estimated as 1~ns ($\sigma$) from prompt positrons
in the incoming beam scattered from the target and hitting the
scintillators.  The time distribution of positrons showed a
prompt peak due to the prompt beam positrons followed by
positrons from muon and muonium decay. The latter contributions
were confirmed by the characteristic spectrum shape determined by
decay lifetime and oscillation of yield due to muon and muonium
spin rotation in the transverse magnetic field.  The prompt beam
positrons were eliminated in the following analysis.

The two-dimensional track extrapolation position distribution
projected along the beam direction consists of three components.
The first, which is dominant in the distribution, comes
from positrons from muon or muonium decay in the target (target
decay). The second is due to positrons from muonium
decay in vacuum (vacuum decay). The third, due to
positrons from decays in other locations such as the support
structure or vacuum chamber walls, became negligible after
applying fiducial volume selections.

\begin{figure}[t]
\begin{center}
\includegraphics[width=0.9\textwidth]
{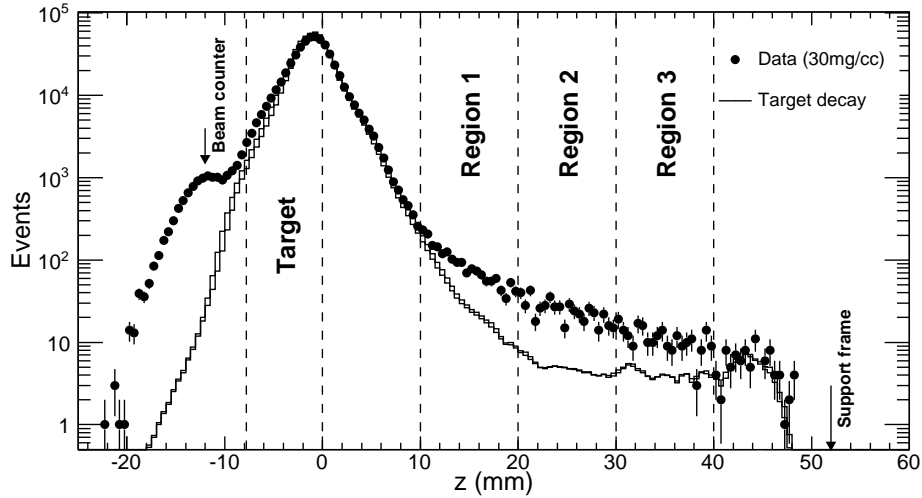}
\caption{\label{F:zdist}
Distribution of positron track extrapolation positions. Closed
circles shows data with 29~mg~cm$^{-3}$ aerogel.
The histogram indicates the target decay distribution estimated from a simulation of muon stopping smeared by the extrapolation resolution.
}
\end{center}
\end{figure}

 The extrapolated spatial distribution for positrons from target decay is described by the convolution of the muon stopping distribution
in the target and the extrapolation resolution. The muon stopping distribution was estimated
with a Monte Carlo simulation by comparison with the measured stopping rate
in the target as a function of beam momentum for each target
density and thickness.
The extrapolated distribution measured with the calibration target was used to characterize 
extrapolation resolution. Figure~\ref{F:zdist} shows the distribution of extrapolated position and 
the estimated contribution from the target-decay positrons. The total number of positron tracks in the region from $z=-$8.4~mm to 40~mm was
used to normalize the target-decay distribution 
to the number of muons stopping in the aerogel target.
This normalization is then independent of the track reconstruction efficiency. 

Positrons from target decay are a dominant contribution near the target surface region, whereas
there is an excess over the target-decay positrons in the vacuum region ($z>10$~mm).

\begin{figure}[t]
\begin{center}
\includegraphics[width=1.0\textwidth]
{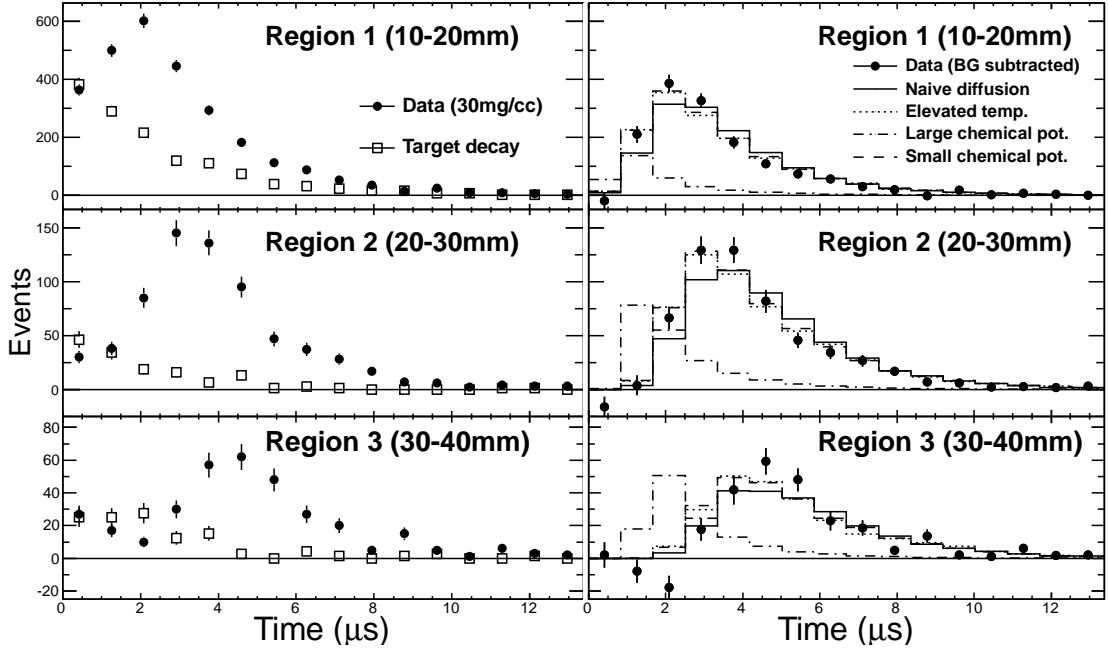}
\caption{\label{F:tdist}
Left plots are time distributions of positrons in each region for
29~mg~cm$^{-3}$ aerogel.
Right plots are background-subtracted time distributions compared with diffusion models.
}
\end{center}
\end{figure}

The time structure of the excess events was examined by dividing
the positron tracks into three $z$ regions.  We define the
regions 1, 2, and 3 as $10<z<20$~mm, $20<z<30$~mm, and
$30<z<40$~mm, respectively.  The time distribution of each region
is shown in Fig.~\ref{F:tdist} for the 29~mg~cm$^{-3}$ aerogel.
In the plots in the left panel, data are shown as closed circles,
and the target-decay events are shown as open squares.
The target-decay events were estimated from data with the calibration
target scaled by a normalization factor that was pre-determined
in the comparison of the $z$-distribution (Fig.~\ref{F:zdist}). The
time distribution of the excess over the target-decay events has a
peak that moves to later time as the selected region moves farther from
the target surface.

The time distributions after subtracting for the target decay events were compared with models of muonium diffusion.
The solid histograms in the right panel of Fig.~\ref{F:tdist} indicate predictions
from three-dimensional diffusion at room temperature (naive diffusion model). In this model,
muonium moves with velocity selected from a Maxwell distribution
from its initial position and diffuses among silica grains
until it decays or is emitted from the target surface. A
parameter in this model is the mean free path between collisions.
The angular distribution of muonium emission that naturally emerges from the target surface is
approximately proportional to $\cos\theta$, where $\theta$ is the polar angle
of muonium with respect to the normal to the surface. 
The naive diffusion model describes basic features of the peaking
structure in time and its evolution with spatial region.
However, it  underestimates data points at early time in region 1 and region 2, which correspond to fast-moving muonium.

Three alternative models were considered to understand the
early-time enhancement compared to naive diffusion. 
First, we consider that muonium is locally heated up when it is formed. Dotted histograms indicate the diffusion model 
with the initial temperature elevated to 400~K.
Another possibility is that the initial muonium energy is
elevated by a chemical potential
as it exits a silica surface within the aerogel sample.
Interpretation of muonium emission to vacuum with a chemical potential as large as 0.3~eV
was reported for data taken with a mesoporous silica film \cite{PSI-Mu}.
We examined whether this could be present in these silica
aerogels, with a simulation represented by dash-dotted histograms in Fig.~\ref{F:tdist}.
We found the simulated early-time enhancement is too large to explain the data for such a large chemical potential (0.3~eV).
However, such a scenario may describe the data if the chemical potential is smaller (25~meV) and
muonium energy is thermally moderated during the diffusion process (small chemical potential model).
This model is shown as the dashed histograms in Fig.~\ref{F:tdist}.
The early-time data points are now better described by either the elevated-temperature or the small chemical-potential 
models.

\begin{table}[t]
\begin{center}
\caption{
\label{T:yieldmfp}
Vacuum yield in the region 1--3 and mean free path}
\begin{tabular}{ c c c l }
\hline
Density (mg~cm$^{-3}$) & Vacuum yield (per 1000 muon stops) &Mean free path ($\mu$m) \\
& & (small chemical potential model)\\
\hline
29   &$2.74 \pm 0.11^{+0.10}_{-0.13}$ &0.226$\pm 0.016^{+0.113}_{-0.079}$\\
47   &$2.81 \pm 0.11^{+0.14}_{-0.08}$ & 0.118$\pm 0.009^{+0.060}_{-0.034}$\\
97   & $3.13 \pm 0.20^{+0.12}_{-0.09}$&0.035$\pm 0.004^{+0.018}_{-0.012}$\\
178 & $1.60 \pm 0.11^{+0.07}_{-0.10}$&0.0050$\pm 0.0007^{+0.0020}_{-0.0014}$ \\
\hline
\end{tabular}
\end{center}
\end{table}

The yield of muonium in the region 1--3 ($10<z<40$~mm) per 1000
muons stopping in the aerogel was found to be
$2.74 \pm 0.11$(stat.)$^{+0.10}_{-0.13}$(syst.) for density 29~mg~cm$^{-3}$. 
This is based only on the background-subtracted data and is thus
independent of the diffusion model.
The mean free path in the diffusion model was estimated from the yield,
assuming the muonium formation fraction of $0.52\pm0.01$ as determined by MuSR.
The mean free path is a model dependent quantity as it depends on the velocity distribution.
The velocity distribution was not uniquely determined from the time distribution
that is a convolution of muonium emission from aerogel and velocity.
The value of the mean free path that best represents the time distributions
varies by 10\% among different models except the large
chemical potential model. 
We obtained 
0.226$\pm 0.016$(stat.)$^{+0.113}_{-0.079}$(syst.) $\mu$m for
density 29~mg~cm$^{-3}$ using the small chemical potential model for the central value.
The systematic uncertainties are dominated by the uncertainty of
the muon beam momentum distribution, which is mostly correlated
for all samples.
Thus, this uncertainty is not relevant in the relative comparison
of densities, except perhaps for the most dense sample. Note that
the vacuum yields of the other three samples are approximately
consistent within the stated uncertainties.
However, comparison with simulations used central beam momenta and stopping distributions
that were unique for each target, where small variations
were important in assessment of mean free paths in the diffusion model for
different densities of aerogel.
The yields and mean free paths for all samples are given in Table~\ref{T:yieldmfp}

\begin{figure}[t]
\begin{center}
\includegraphics[width=0.7\textwidth]
{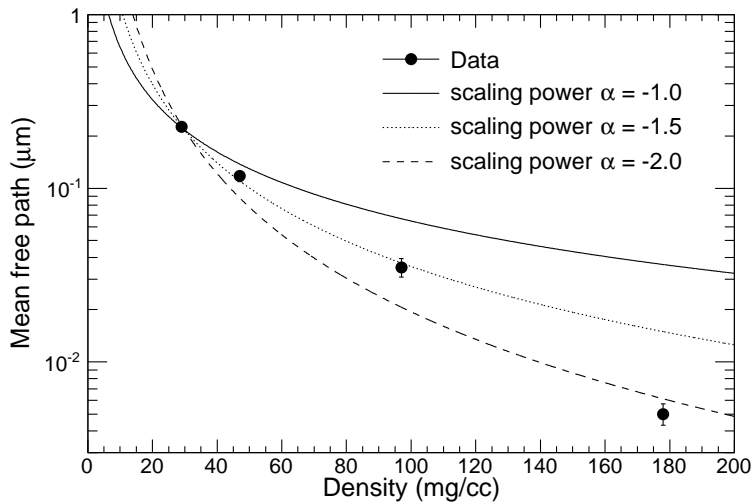}
\caption{\label{F:mfp}
Mean free path and density of silica aerogel.
}
\end{center}
\end{figure}

Density dependence of the mean free path can be parameterized using a scaling power relation 
$L=L_0(\frac{\rho}{\rho_0})^{\alpha}$ where $\rho$ is the density,
 $\rho_0$ and $L_0$ are  reference density (29 mg~cm$^{-3}$) and its mean free path,
 and $\alpha$ is a parameter.
It is known that the muon stopping density (stopping muons per unit sample
thickness) scales with the
material density as $\alpha_{\mu}=+1.0$, i.e., there is a linear
proportionality between stopping density and material density.
The total yield in the vacuum region should scale as $\alpha/2 +
\alpha_{\mu}$ since the mean one-dimensional distance travelled
in the diffusion process scales
as $\sqrt{L}$ for isotropic scattering.
The value of $\alpha$ is predicted to be $-$1.0 
in a simple geometric model that assumes silica aerogel consisting of grains of SiO$_2$ spheres.
In this case, higher yield is expected at higher density, but
this gain with density would disappear if $\alpha<-2.0$.
The density dependence of the mean free path is shown in  Fig~\ref{F:mfp},
where data are compared with three values of $\alpha$.
The measured dependence is clearly larger than for $\alpha=-1.0$;
the dotted curve with $\alpha=-1.5$ best describes the data,
except for the 178~mg~cm$^{-3}$ aerogel.
This would imply that the underlying mechanism for the density
dependence is beyond the simple geometric picture.

\begin{table}[t]
\begin{center}
\caption{
\label{T:BET}
 Specific surface area and estimated grain radius }
\begin{tabular}{ c c c }
\hline
Density (mg~cm$^{-3}$) & Specific surface area (m$^2$\,g$^{-1}$) & Estimated grain radius (nm) \\
\hline
29   &$550$ & 2.5\\
47   &$620$ & 2.2\\
97   &$716$ & 1.9\\
178 & $914$ & 1.5\\
\hline
\end{tabular}
\end{center}
\end{table}

The BET method \cite{BET} was used to estimate the specific
surface area of all samples.
The grain radius was estimated by assuming spherical shapes of
SiO$_2$ grains in the aerogel structure. 
They are shown in Table~\ref{T:BET}.
It suggests that the grain size deduced from an assumption 
of SiO$_2$ nano-spheres decreases with increasing sample density. 
Images from an electron-microscope (SEM/TEM)
showed a grain-network structure, but the density dependence is
difficult to interpret and is not considered in this discussion.
These microscopic differences may play an important role behind the observed density dependence.

\section{Discussion and prospects}
For the comparison with the application of the $g-2$ experiment at J-PARC,
the measured yield of vacuum emission needs to be scaled to match the beam conditions.
According to the naive diffusion model discussed above, 
the total yield of Mu in vacuum $y_{vac}$ from 29 mg~cm$^{-3}$ is
$4.8 \times 10^{-3}$ per incident muon 
in a beam with a momentum spread of 2\% (RMS) at 23~MeV/c.
This includes those decaying outside of our
measurement fiducial regions.
The beam momentum and its spread at J-PARC is designed to be 28~MeV/c and 5\% (RMS), respectively.
Under the assumption that only a small region near the surface
contributes to emission, the yield is proportional to
the muon stopping density $\rho_{stop}$ at the surface, which in
turn is inversely 
proportional to the range spread $\Delta R$ of the muon stopping distribution.
For a surface muon beam, 
assuming the relatively small range straggling term~\cite{Pifer}
in the range spread can be ignored,
the range spread is proportional to the total range $R$.
Total range $R$ scales with momentum as $p^{3.5}$
with the proportionality constant depending on the the relative
momentum spread of the beam $\delta p=\frac{\Delta p}{p}$.
Therefore, the vacuum yield $\tilde{y}_{vac}$ with other beam conditions is estimated from $y_{vac}$ as

\begin{eqnarray}
 \tilde {y}_{vac} 
 = \left( \frac{ \tilde{ \rho }_{stop}  } { \rho_{stop} } \right) y_{vac}
 = \left( \frac{ \Delta R } { \Delta \tilde{R} } \right) y_{vac}
 			 =  \left( \frac{\delta p}{ \delta
                             \tilde{ p } } \right)
                         \left(\frac{p}{ \tilde{p} }\right)^{3.5}
                         y_{vac} ,
\end{eqnarray}
where variables with a tilde indicate conditions to be estimated.
In the case of J-PARC beam conditions, we obtained
$\tilde{y}_{vac} = 9.6 \times 10^{-4}$.

The probability of muonium emission to vacuum is calculated by
using diffusion models with parameters 
estimated in this work. In the models, the probability drops
exponentially as a function of 
distance of the initial muonium formation point from the surface .
For 29~mg~cm$^{-3}$ aerogel, the probability goes down to $e^{-1}$ at
30~$\mu$m from the surface in the small chemical potential model.
On the other hand, the stopping distribution of incoming muons spreads over a few millimeters due to the energy
loss and range straggling and the momentum width of the muon beam.
Therefore the length scale of diffusion is much shorter than that of 
the stopping distribution. This is one of the major limitations to higher muonium production yield in vacuum.
The consequence of this is that improvement of yield in vacuum
might be anticipated
if one introduces an intermediate structure which is porous or
has peaks and valleys
of 100~$\mu$m in scale so that muonium may reach the surface promptly
during the diffusion process before decay. Development of such a silica aerogel
sample will be the subject for future research to increase the yield.

\section{Conclusions}
Muonium emission from several silica aerogel samples of different
densities was confirmed in this work.
The space-time distribution of muonium follows a naive diffusion model.
Some possible deviations from this naive diffusion model in the early part of the time spectrum are observed.
The density dependence of the mean free path is weaker than
predicted by naive scaling, and there is no strong enhancement of
the yield at higher densities.
There is potential for an increase of the production yield by introduction of intermediate structure having a scale of 
approximately 100~$\mu$m.

\ack

The authors are pleased to acknowledge the support  
from TRIUMF to provide a stable beam
during the experiment. Special thanks go to R. Henderson,
R. Openshaw, G. Sheffer, and M. Goyette from the TRIUMF Detector
Facility. We also thank D. Arseneau,
G. Morris, B. Hitti, R. Abasalti, and D. Vyas of the TRIUMF
Materials and Molecular Science Facility.
Research was supported in part by the 
MEXT KAKENHI Grant Number 231800N (Japan) and NSERC Discovery Grant (Canada).

\end{document}

%% file: authors.tex
\author{
 \name{P.~Bakule}{9},
 \name{G.A.~Beer}{3},
 \name{D.~Contreras}{10},
 \name{M.~Esashi}{1},
 \name{Y.~Fujiwara}{8,5},
 \name{Y.~Fukao}{7},
 \name{S.~Hirota}{7,5},
 \name{H.~Iinuma}{7},
 \name{K.~Ishida}{8},
 \name{M.~Iwasaki}{8},
 \name{T.~Kakurai}{8,5},
 \name{S.~Kanda}{8,5},
 \name{H.~Kawai}{4},
 \name{N.~Kawamura}{7},
 \name{G.M.~Marshall}{10},
 \name{H.~Masuda}{2},
 \name{Y.~Matsuda}{6},
 \name{T.~Mibe}{7},
 \name{Y.~Miyake}{7},
 \name{S.~Okada}{8},
 \name{K.~Olchanski}{10},
 \name{A.~Olin}{10,3},
 \name{H.~Onishi}{8},
 \name{N.~Saito}{7,5},
 \name{K.~Shimomura}{7},
 \name{P.~Strasser}{7},
 \name{M.~Tabata}{4},
 \name{D.~Tomono}{8,
 \thanks{Present Address: School of Physics and Astronomy, Queen Mary University of London, Mile End
Road, London E1 4NS, UK}},
 \name{K.~Ueno}{7},
 \name{K.~Yokoyama}{8,
 \thanks{Present Address: Dept. of Physics, Kyoto University, Kyoto, Japan}},
 \name{S.~Yoshida}{1}
}

\address{
 \affil{1}{Advanced Institute for Materials Research, Tohoku University, Sendai 980-8578,Japan}
 \affil{2}{Division of Applied Chemistry, Tokyo Metropolitan University, Tokyo, 192-0397, Japan}
 \affil{3}{Department of Physics and Astronomy, University of Victoria, Victoria BC V8W 3P6, Canada}
 \affil{4}{Department of Physics, Chiba University, Chiba 263-8522, Japan}
 \affil{5}{Department of Physics, The University of Tokyo, Tokyo, 113-0033, Japan}
 \affil{6}{Graduate School of Arts and Sciences, The University of Tokyo, Tokyo, 153-8902, Japan}
 \affil{7}{High Energy Accelerator Research Organization (KEK), Ibaraki, 305-0801, Japan}
 \affil{8}{RIKEN Nishina Center, RIKEN, Saitama, 351-0198, Japan}
 \affil{9}{RIKEN–RAL Muon Facility, Rutherford Appleton Laboratory, Harwell Oxford, Didcot, Oxfordshire, OX11 0QX, UK}
 \affil{10}{TRIUMF, Vancouver, BC, V6T 2A3, Canada}
}